\documentclass[10pt, conference]{IEEEtran}

\usepackage{amsmath}

\usepackage{cite, url}
\usepackage{graphicx}
\usepackage{epstopdf}
\usepackage[subrefformat=parens, labelformat=parens]{subfig}
\usepackage[font=footnotesize, labelfont=footnotesize]{caption}
\usepackage{color}
\usepackage{lipsum}
\usepackage{enumitem}
\usepackage{listings}
\usepackage{hyperref}

\makeatletter
\AtBeginDocument{%
  \let\c@figure\c@lstlisting
  
  \let\ftype@lstlisting\ftype@figure 
}
\makeatother

\usepackage[keeplastbox]{flushend}

\usepackage{siunitx} 
\DeclareSIUnit \MHz{ MHz }
\DeclareSIUnit \GHz{ GHz }
\DeclareSIUnit \dBm{ dBm }
\DeclareSIUnit \dB{ dB}

\newcommand{\mytexttt}[1]{\texttt{\small #1}}

\begin{document}

\sloppy

\title{Open-Access Full-Duplex Wireless \\ in the ORBIT Testbed}

\author{
Tingjun Chen\IEEEauthorrefmark{1}, Mahmood Baraani Dastjerdi\IEEEauthorrefmark{1}, Guy Farkash\IEEEauthorrefmark{1}, Jin Zhou\IEEEauthorrefmark{2}, Harish Krishnaswamy\IEEEauthorrefmark{1}, Gil Zussman\IEEEauthorrefmark{1} \\
\IEEEauthorrefmark{1}Electrical Engineering, Columbia University, New York, NY 10027, USA \\
\IEEEauthorrefmark{2}Electrical and Computer Engineering, University of Illinois at Urbana-Champaign, Urbana, IL 61801, USA \\
\{tingjun@ee., b.mahmood@, guy.farkash@, harish@ee., gil@ee.\}columbia.edu, jinzhou@illinois.edu
}

\maketitle

\setlength{\abovedisplayskip}{3pt}
\setlength{\belowdisplayskip}{3pt}

\begin{abstract}
In order to support experimentation with full-duplex (FD) wireless, we recently integrated an open-access FD transceiver in the ORBIT testbed~\cite{orbit}. In this report, we present the design and implementation of the FD transceiver and interfaces, and provide examples and guidelines for experimentation. In particular, an ORBIT node with a National Instruments (NI)/Ettus Research Universal Software Radio Peripheral (USRP) N210 software-defined radio (SDR) was equipped with the Columbia FlexICoN Gen-1 customized RF self-interference (SI) canceller box. The RF canceller box includes an RF SI canceller that is implemented using discrete components on a printed circuit board (PCB) and achieves $\SI{40}{dB}$ RF SI cancellation across $\SI{5}{MHz}$ bandwidth.
We provide an FD transceiver baseline program and present two example FD experiments where $\SI{90}{dB}$ and $\SI{85}{dB}$ overall SI cancellation is achieved for a simple waveform and PSK modulated signals across both the RF and digital domains.
We also discuss potential FD wireless experiments that can be conducted based on the implemented open-access FD transceiver and baseline program.
\end{abstract}

\section{Introduction}
\label{sec:intro}
Due to its potential to double network capacity at the physical (PHY) layer and to provide many other benefits at higher layers, full-duplex (FD) wireless has drawn significant attention~\cite{sabharwal2014band,bharadia2013full,zhou2017integrated}. The major challenge associated with FD is the extremely strong self-interference (SI) on top of the desired signal, requiring more than $\SI{90}{dB}$ of self-interference cancellation (SIC) across both the RF and digital domains.

Our work on FD transceivers/systems within the Columbia FlexICoN project~\cite{flexicon} focuses on integrated circuit (IC) implementations that are appropriate for mobile and small-form-factor devices~\cite{zhou2017integrated,Zhou_NCSIC_JSSC14,krishnaswamy2016full}.
In~\cite{fd_demo_mobihoc16}, we presented the FlexICoN Gen-1 FD transceiver and an FD wireless link, featuring $\SI{40}{dB}$ RF SIC across $\SI{5}{MHz}$. The implemented Gen-1 RF SI canceller emulates its RFIC counterpart that we presented in~\cite{Zhou_NCSIC_JSSC14} and modeled and analyzed in~\cite{marasevic2017resource}.

However, there is no existing open-access wireless testbed with FD-capable nodes, which is crucial for experimental evaluations of FD-related algorithms at the higher layers. Therefore, to facilitate research in this area and to allow the broader community to experiment with FD wireless, we integrated an improved version of the Gen-1 RF canceller presented in~\cite{fd_demo_mobihoc16} with a National Instruments (NI)/Ettus Research USRP N210 SDR in the open-access ORBIT wireless testbed~\cite{orbit}. Since interfacing an RFIC canceller with an SDR presents numerous technical challenges, we implemented the RF canceller on a printed circuit board (PCB) to facilitate the cross-layered experiments with an SDR platform.

In this technical report, we present our cross-layered (hardware and software) implementation of an open-access FD transceiver integrated with the ORBIT testbed, including the design and implementation of the customized Gen-1 RF canceller box and an FD transceiver baseline program. We also present two example FD experiments that run remotely in the ORBIT testbed, where SIC is performed across both the RF and digital domains, demonstrating the FD capability in the ORBIT testbed. The first example is based on UHD~\cite{uhd}, where $\SI{90}{dB}$ overall SIC is achieved for a simple waveform. The second example is based on GNU Radio~\cite{gnuradio}, where $\SI{85}{dB}$ overall SIC is achieved for PSK modulated signals. The code for the baseline program and a tutorial for the FD transceiver are available at~\cite{flexicon_github,flexicon_orbit_gen1}. The implemented FD transceiver and the baseline program, which can be further extended to more complicated communication networking scenarios, can allow the broader community to experiment with FD wireless.

\section{The FlexICoN Gen-1 RF Canceller Box}
\label{sec:canceller}

\begin{figure}[!t]
\centering
\includegraphics[width=0.98\columnwidth]{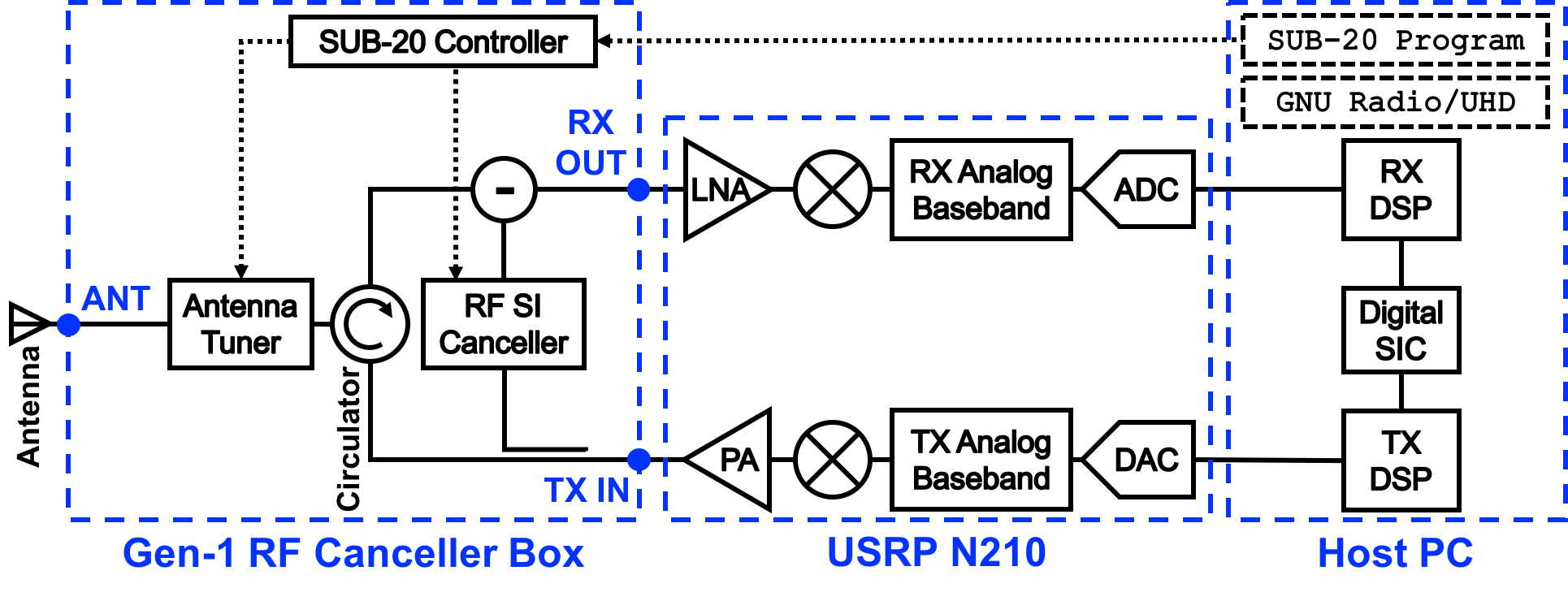}
\caption{Block diagram of the implemented FD transceiver.}
\label{fig:orbit-node-diagram}
\vspace{-0.5\baselineskip}
\end{figure}
\begin{figure}[!t]
\centering
\subfloat[]{
\label{fig:orbit-node-canc}
\includegraphics[width=0.49\columnwidth]{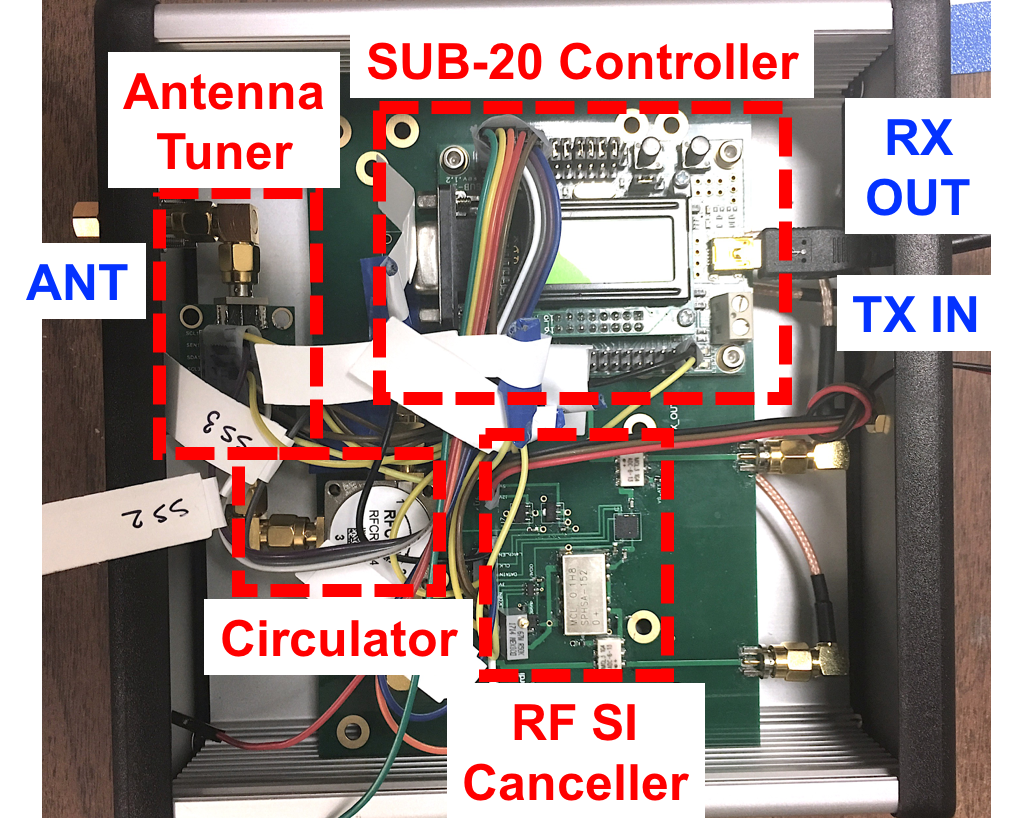}}
\hfill
\subfloat[]{
\label{fig:orbit-node-sdr}
\includegraphics[width=0.465\columnwidth]{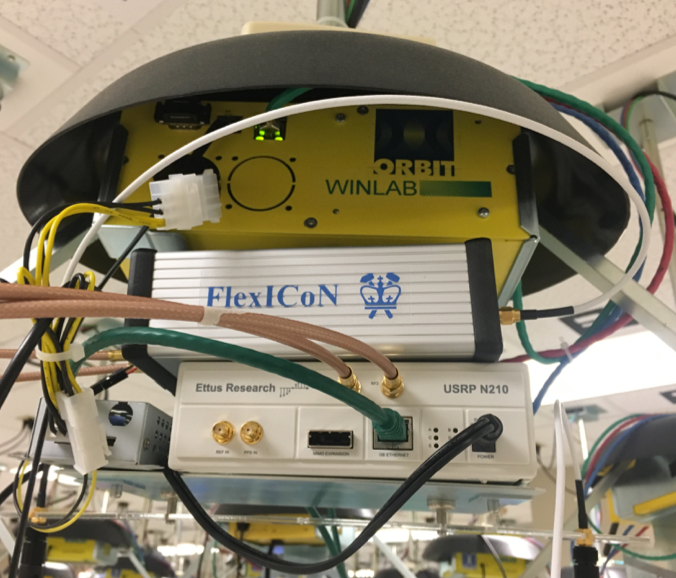}}
\caption{(a) The Columbia FlexICoN Gen-1 RF canceller box, and (b) the FD-capable node installed in the ORBIT wireless testbed.}
\label{fig:orbit-node}
\vspace{-0.5\baselineskip}
\end{figure}

Fig.~\ref{fig:orbit-node-diagram} shows the block diagram of the implemented FD transceiver, in which a Gen-1 RF canceller box (as depicted in Fig.~\ref{fig:orbit-node}\subref{fig:orbit-node-canc}) is connected to an Apex II multi-band antenna (at the ANT port) and a USRP (at the TX IN and RX OUT ports). Fig.~\ref{fig:orbit-node}\subref{fig:orbit-node-sdr} shows the FD transceiver installed in the ORBIT testbed. Specifically, a circulator is used at the antenna interface so that a single antenna can be shared between the TX and RX. To alleviate the RX front-end linearity and the analog-to-digital converter (ADC) dynamic range requirements, sufficient SI isolation and cancellation in the RF domain are needed before digital SIC is engaged.

\begin{figure}[!t]
\centering
\includegraphics[width=0.5\columnwidth]{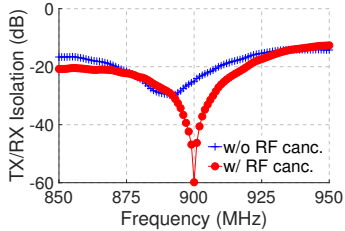}
\caption{Measured TX/RX isolation of the RF canceller box with and without turning on the RF canceller. The RF canceller box with the circulator and the RF canceller provides $\SI{40}{dB}$ RF SIC across $\SI{5}{MHz}$ bandwidth.}
\label{fig:meas-canc-box}
\vspace{-0.5\baselineskip}
\end{figure}

In the FD transceiver, the RF SI suppression is achieved by the circulator and the RF SI canceller in the Gen-1 RF canceller box, where the circulator has a TX/RX isolation of around $\SI{20}{dB}$ and the RF SI canceller can provide $20$-$\SI{30}{dB}$ RF SIC. As Fig.~\ref{fig:orbit-node}\subref{fig:orbit-node-canc} shows, the RF canceller box contains four components: (i) a frequency-flat amplitude- and phase-based RF canceller, which is an improved version of that presented in~\cite{fd_demo_mobihoc16}\footnote{The implemented RF canceller includes a variable gain attenuator with higher resolution and an SPI compared with that presented in~\cite{fd_demo_mobihoc16}.}, (ii) a coaxial circulator, (iii) a custom-designed antenna tuner, and (iv) a SUB-20 controller. Fig.~\ref{fig:meas-canc-box} shows an example of the measured TX/RX isolation (measured between TX IN and RX OUT ports of the canceller box), where $\SI{40}{dB}$ RF SIC is achieved across $\SI{5}{MHz}$ bandwidth. 

\begin{figure}[t]
\centering
\subfloat{
\label{fig:meas-canc-no-circ-amp}
\includegraphics[width=0.48\columnwidth]{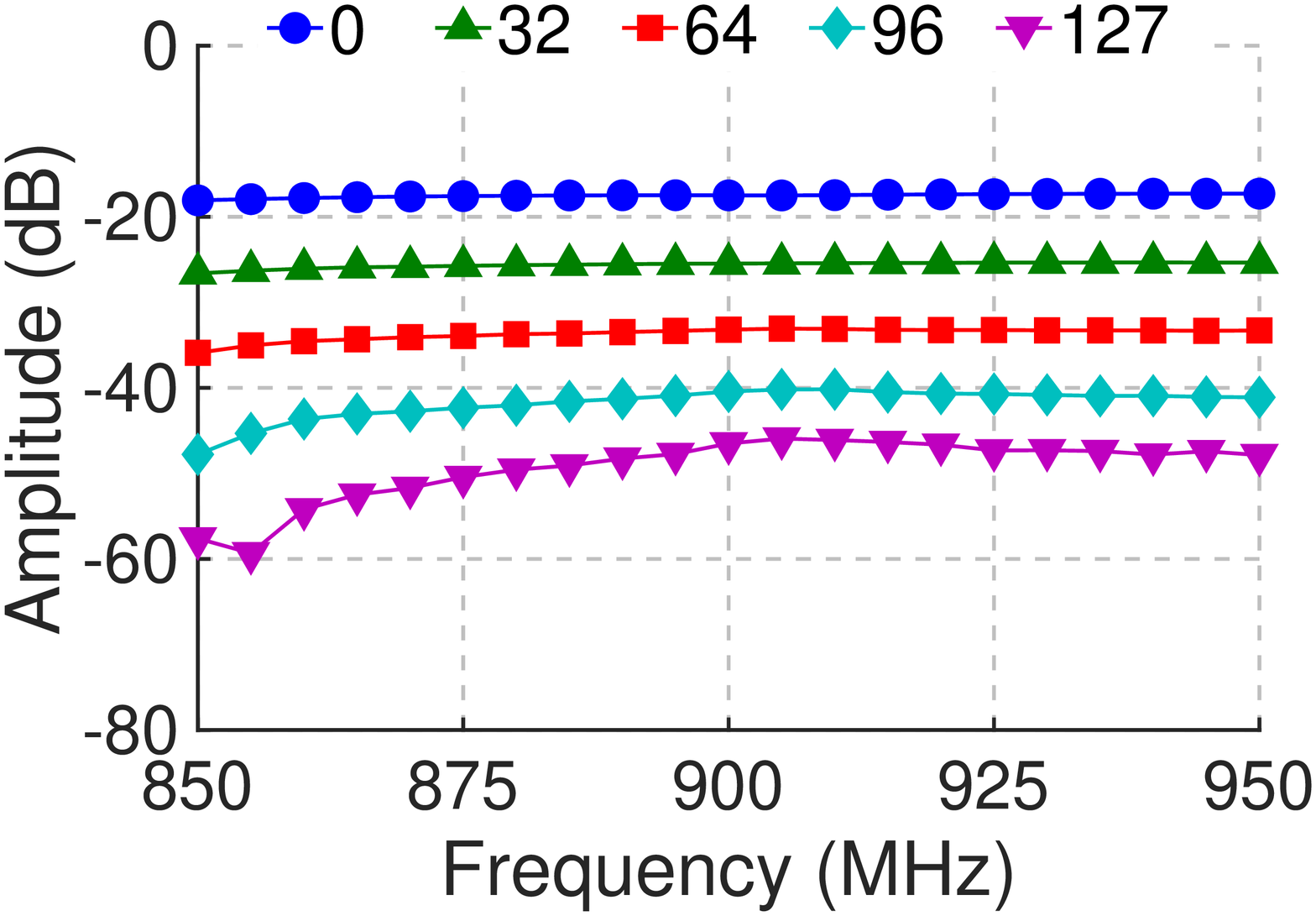}}
\hspace{-12pt} \hfill
\subfloat{
\label{fig:meas-canc-no-circ-phase}
\includegraphics[width=0.48\columnwidth]{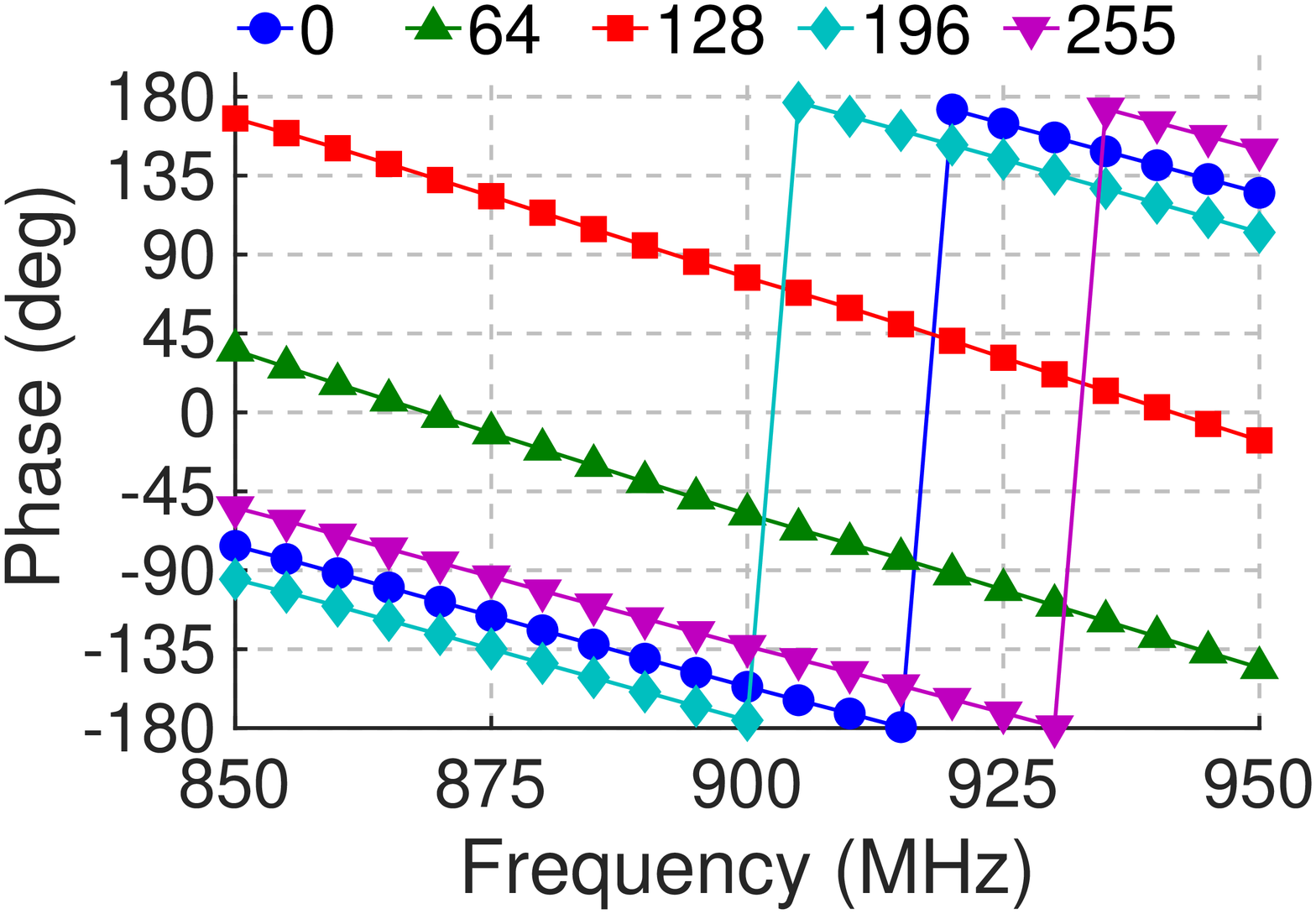}}
\caption{Measured amplitude and phase of the RF canceller with varying attenuation \mytexttt{ATT} values (left) and phase shift \mytexttt{PS} values (right).}
\label{fig:meas-canc-no-circ}
\vspace{-0.5\baselineskip}
\end{figure}

\subsection{The Amplitude- and Phase-based RF Canceller}
The amplitude- and phase-based RF canceller is implemented using discrete components on a PCB and is optimized around $\SI{900}{MHz}$ operating frequency.\footnote{In this implementation, we select $\SI{900}{MHz}$ operating frequency but this approach can be easily extended to other frequencies (e.g., $2.4/\SI{5}{GHz}$).} The RF canceller taps a reference signal from the output of the power amplifier (PA) at the TX side (through a $\SI{6}{dB}$ Mini-Circuits ADC-6-13+ directional coupler) and adjusts its amplitude and phase. Then, SIC is performed at the input of the low-noise amplifier (LNA) at the RX side.

For amplitude adjustment, a $7$-bit SKY12343-364LF digital attenuator~\cite{SKY12343} is used, in which the attenuation can be adjusted within a $\SI{31.75}{dB}$ range with a resolution of $\SI{0.25}{dB}$. As a result, the RF canceller has an amplitude tuning range between $-\SI{48}{dB}$ and $-\SI{17}{dB}$. For phase adjustment, a Mini-Circuits passive SPHSA-152+ phase-shifter~\cite{SPHSA152} is used, which covers full $\SI{360}{deg}$ and is controlled by an $8$-bit TI-DAC081S101 digital-to-analog converter (DAC)~\cite{DAC081S101}. Both the attenuator and phase shifter are programmed through the SUB-20 controller serial-to-parallel interface (SPI) with code values \mytexttt{ATT} (\underline{ATT}uation) and \mytexttt{PS} (\underline{P}hase \underline{S}hift), respectively, and the parameter configuration ranges are
\begin{align*}
\mytexttt{ATT} \in \{0,1,\cdots,127\},\ \mytexttt{PS} \in \{0,1,\cdots,255\}.
\end{align*}
The attenuator and DAC have $\SI{3}{V}$ supply voltage and the phase shifter has a reference voltage of $\SI{12}{V}$.

Fig.~\ref{fig:meas-canc-no-circ} shows the amplitude and phase measurements of the RF canceller with varying \mytexttt{ATT} values (under fixed $\mytexttt{PS} = 0$) and with varying \mytexttt{PS} values (under fixed $\mytexttt{ATT} = 0$). As Fig.~\ref{fig:meas-canc-no-circ} shows, the RF canceller has an amplitude tuning range of $\SI{29}{dB}$ (from $-\SI{46.5}{dB}$ to $-\SI{17.5}{dB}$) and a phase tuning range of full $\SI{360}{deg}$.

\subsection{The Coaxial Circulator}
An RF-CI RFCR3204 coaxial circulator is used, whose operating frequency is between $860$-$\SI{960}{MHz}$.


\begin{figure}[t]
\centering
\includegraphics[width=0.95\columnwidth]{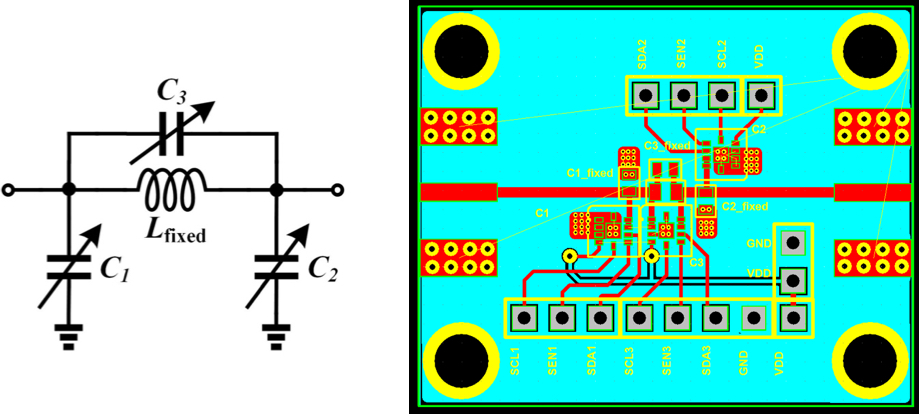}
\caption{Circuit diagram and PCB implementation of the programmable antenna tuner.}
\label{fig:tuner}
\vspace{-0.5\baselineskip}
\end{figure}

\subsection{The Programmable Antenna Tuner}
In order for the circulator to better match with varying impedance of the antenna due to environmental changes (around $\SI{900}{MHz}$ operating frequency), we also designed and implemented a programmable antenna tuner. Fig.~\ref{fig:tuner} shows the circuit diagram and the PCB implementation of the antenna tuner. In particular, a $\pi$-network with lossless inductor ($L$) and digitally tunable capacitors ($C_i$) is used for impedance transformation. In our implementation, we use a fixed chip inductor with inductance $L_{\rm fixed} = \SI{5.1}{nH}$ and the Peregrine Semiconductor $5$-bit PE64909 digitally tunable capacitors~\cite{PE64909} for $C_i$ ($i=1,2,3$). By programming the capacitors with code values \mytexttt{CAPi} ($i=1,2,3$), different antenna interface impedance matching can be achieved. The corresponding configuration ranges of the tunable capacitors are
\begin{align*}
\mytexttt{CAPi} \in \{0,1,\cdots,31\},\ \forall i = 1,2,3.
\end{align*}

\subsection{The SUB-20 Controller}
As Fig.~\ref{fig:orbit-node-diagram} shows, a DIMAX SUB-20 multi-interface USB adapter~\cite{SUB20} connected to the host PC is used to program the attenuator and DAC (on the RF SI canceller) and the capacitors (on the antenna tuner) through SPI. The SUB-20 SPI is configured to operate at the maximal master clock of $\SI{8}{MHz}$. At this clock rate, programming one \mytexttt{ATT} or \mytexttt{PS} value (a $2$-byte word including the address fields, etc.) takes $\SI{2}{us}$, and programming one \mytexttt{CAPi} value (a $1$-byte word) takes $\SI{1}{us}$. We note that other controller platforms with higher SPI clock rates can also be used to improve the performance.

\section{Integration with the ORBIT Testbed \\ and an FD Transceiver Baseline Node Image}
\label{sec:integration}
An ORBIT node equipped with the Gen-1 RF canceller box is depicted in Fig.~\ref{fig:orbit-node}\subref{fig:orbit-node-sdr}. We use \mytexttt{node11-10} in the ORBIT main \mytexttt{grid} with a USRP N210 SDR. In particular, the RF canceller box TX IN/RX OUT ports are connected to the USRP TX/RX ports, respectively, and the RF canceller box ANT port is connected to an Apex II multi-band antenna (see Figs.~\ref{fig:orbit-node-diagram} and~\ref{fig:orbit-node}).
We developed an FD transceiver baseline node image, which contains two example FD experiments running on the host PC (i.e., the yellow box in Fig.~\ref{fig:orbit-node}\subref{fig:orbit-node-sdr}): (i) a UHD-based example with a simple waveform, and (ii) a GNU Radio-based example with modulated signals using Phase-Shift Keying (PSK) modulation scheme. Throughout the experiments, the USRP has a receiver noise floor of $-\SI{85}{dBm}$.\footnote{This USRP receiver noise floor is limited by the existence of environmental interference at $\SI{900}{MHz}$ frequency. The USRP has a true noise floor of around $-\SI{95}{dBm}$ at the same receiver gain setting, when not connected to an antenna.}

To facilitate the experiments with the RF canceller box and FD wireless, the customized FD transceiver baseline node image named \mytexttt{flexicon-orbit-v2.ndz} with the required software was created and stored in the ORBIT testbed. The code for the FD transceiver baseline program is available at \href{https://github.com/Wimnet/flexicon_orbit}{\mytexttt{https://github.com/Wimnet/flexicon\_orbit}}. The detailed tutorial and instructions containing the steps for running the example FD experiments can be found at~\cite{flexicon_orbit_gen1, flexicon_github}.

\section{An Example FD Experiment based on UHD}
\label{sec:exp-simple}

\begin{lstlisting}[float=tp,basicstyle=\ttfamily\footnotesize,caption={Representative output of the FD transceiver SUB-20 \mytexttt{C} program.},label={fig:orbit-output-sub20},captionpos=b,language=c++,frame=single,belowskip=-.5\baselineskip,linewidth=0.96\columnwidth,xleftmargin=0.02\columnwidth]
$ ./rf_canc_gen1_config 30 110 16 6 6
Sub20 device found... Device opened!
Finished programming ATT with value 30
Finished programming PS with value 110
Finished programming CAP1 with value 16
Finished programming CAP2 with value 6
Finished programming CAP3 with value 6
\end{lstlisting}
\begin{lstlisting}[float=tp,basicstyle=\ttfamily\footnotesize,caption={Representative output of the FD transceiver UHD program.},label={fig:orbit-output-sic},captionpos=b,language=c++,frame=single,belowskip=-.5\baselineskip,linewidth=0.96\columnwidth,xleftmargin=0.02\columnwidth]
$ ./fd_transceiver_simple --rate 5e6 --freq 900e6
 --tx-gain 10 --rx-gain 10 --wave-freq 200e3
...
TX Signal: 0.00 dBm
RX Signal after RF SIC: -45.21 dBm
Amount of RF SIC: 45.21 dB
RX Signal after Digital SIC: -87.87 dBm
Amount of Digital SIC: 42.66 dB

TX Signal: 0.00 dBm
RX Signal after RF SIC: -45.28 dBm
Amount of RF SIC: 45.28 dB
RX Signal after Digital SIC: -88.53 dBm
Amount of Digital SIC: 43.25 dB
...
\end{lstlisting}
\begin{figure}[!t]
\centering
\vspace{-.5\baselineskip}
\subfloat[]{
\label{fig:exp-simple-node}
\includegraphics[width=0.48\columnwidth]{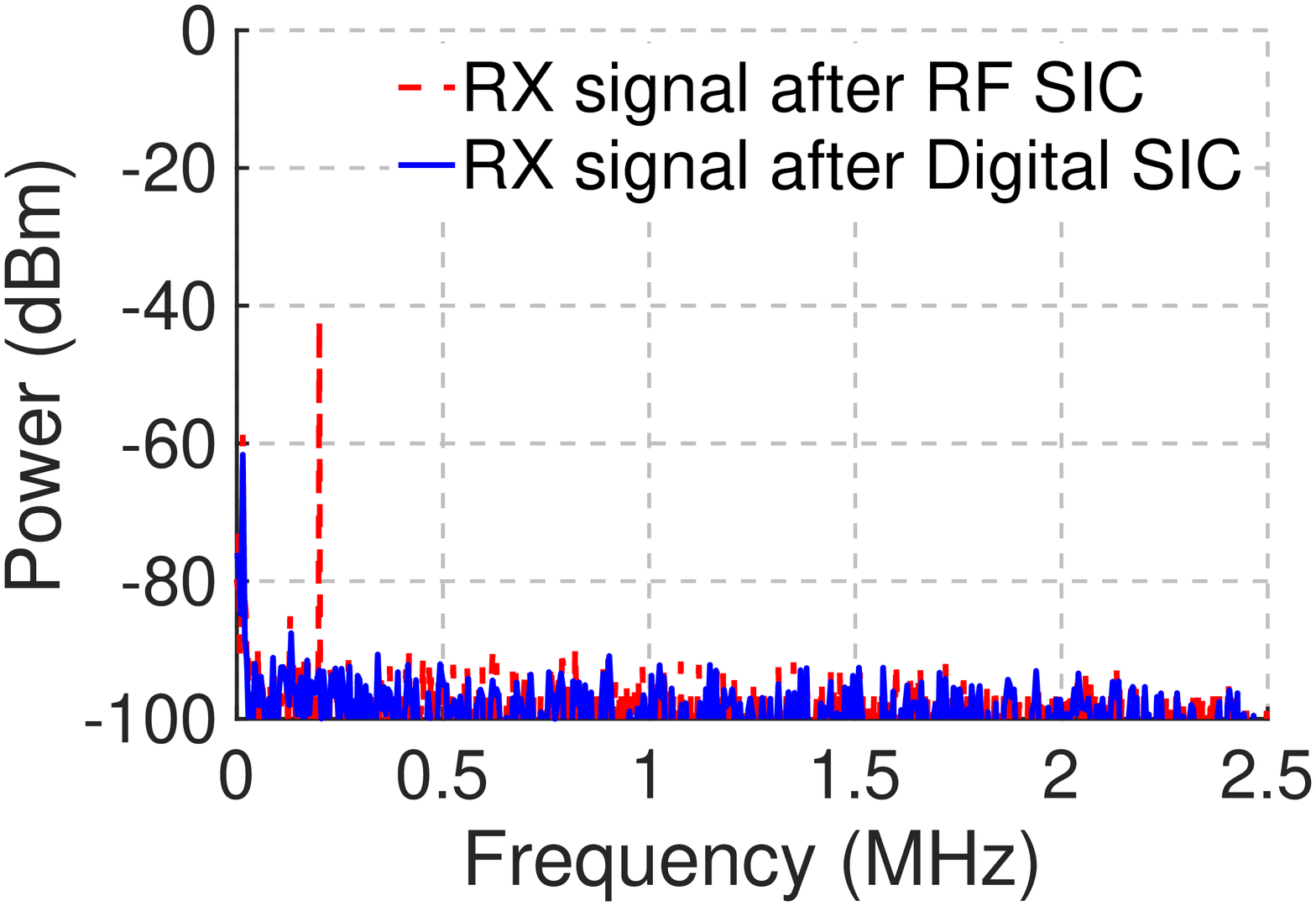}}
\hspace{-12pt} \hfill
\subfloat[]{
\label{fig:exp-simple-link}
\includegraphics[width=0.48\columnwidth]{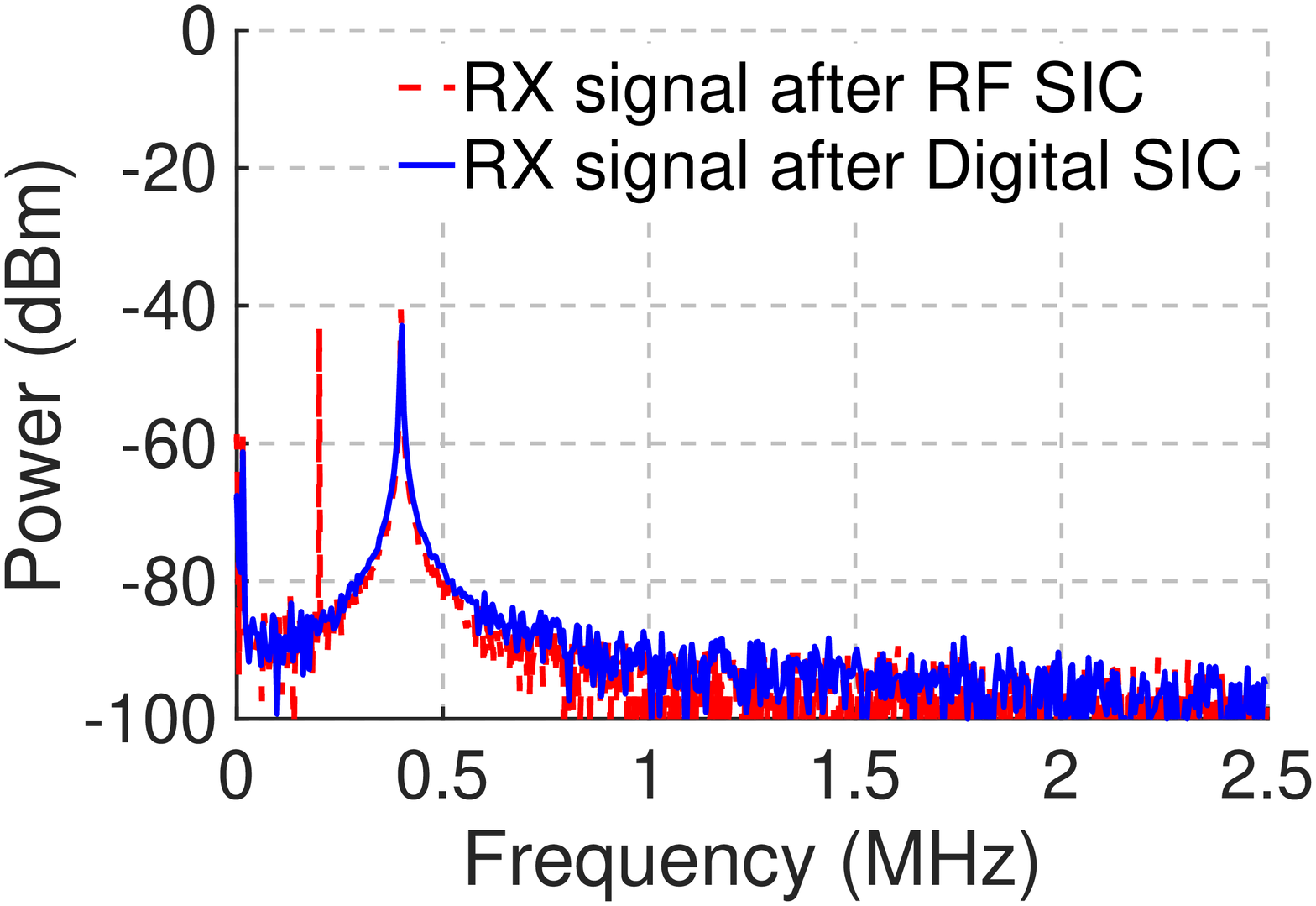}}
\vspace{-0.5\baselineskip}
\caption{Power spectrum of the received signal at the FD transceiver at $\SI{0}{dBm}$ TX power: (a) without the desired signal, (b) with the desired signal.}
\label{fig:exp-simple}
\vspace{-0.5\baselineskip}
\end{figure}

In this section, we present an example FD experiment using the FD transceiver and the baseline program, where the FD transceiver transmits and receives simultaneously at $\SI{900}{MHz}$ carrier frequency with $\SI{5}{MHz}$ sampling rate. Different from regular UHD programs that are designed for half-duplex applications, the FD UHD program includes three parallel threads for performance optimization: the TX/RX streaming threads running on the same frequency channel and a third thread for executing the digital SIC algorithm. In particular, the digital SIC algorithm is based on Volterra series and a least-square problem and is similar to that presented in~\cite{bharadia2013full,krishnaswamy2016full}. Moreover, the \mytexttt{Eigen C++} library is included for computations in the digital SIC algorithm (e.g., matrix operations and FFT).

In this example FD experiment, the FD transceiver (\mytexttt{node11-10}) sends a single tone with frequency offset $\SI{200}{kHz}$ at $\SI{5}{dBm}$ TX power level. Fig.~\ref{fig:orbit-output-sub20} shows an example output of the FD transceiver SUB-20 program, where the RF canceller box is configured with parameters
\begin{align*}
\mytexttt{ATT} = 30,\ \mytexttt{PS} = 110,\ \mytexttt{CAP1} = 16,\ \mytexttt{CAP2} = 6,\ \mytexttt{CAP3} = 6,
\end{align*}
through the \mytexttt{C} program \mytexttt{rf\_canc\_gen1\_config}.\footnote{The optimal configuration of the RF canceller box may change due to factors such as antenna being re-tightened or rotated. Please refer to the detailed tutorial~\cite{flexicon_orbit_gen1} for updates.} Fig.~\ref{fig:meas-canc-box} shows the TX/RX isolation of the RF canceller box under this configuration. Fig.~\ref{fig:orbit-output-sic} shows an example output of the FD transceiver UHD program where $\SI{90}{dB}$ overall SIC is achieved, where $\SI{45}{dB}$ is from the RF domain and $\SI{45}{dB}$ is from the digital SIC algorithm, and the SI signal is canceled to the receiver noise floor. Fig.~\ref{fig:exp-simple} shows the power spectrum of the residual SI after RF and digital SIC through an offline MATLAB script.

In addition, another ORBIT node (\mytexttt{node13-8}) serves as a second radio that sends a single tone with frequency offset $\SI{400}{kHz}$ using the UHD \mytexttt{tx\_waveforms} program~\cite{uhd}, i.e.,
\begin{lstlisting}[basicstyle=\ttfamily\small,language=c++]
$ ./tx_waveforms --rate 5e6 --freq 900e6
 --wave-type SINE --wave-freq 400e3
\end{lstlisting}
Fig.~\ref{fig:exp-simple} presents the power spectrum of the signal received at the FD transceiver after RF and digital SIC. As Fig.~\ref{fig:exp-simple} shows, the SI at the FD transceiver (with frequency offset $\SI{200}{kHz}$) is canceled to the receiver noise floor after SIC in both the RF and digital domains, and the digital SIC algorithm introduces minimal SNR loss to the desired signal (with frequency offset $\SI{400}{kHz}$).

\section{An Example FD Experiment based on GNU Radio}
\label{sec:exp-psk}
In this section, we present another example FD experiment based on GNU Radio, where the FD transceiver transmits a wideband PSK-modulated signals. Compared with the UHD-based example, GNU Radio provides both user-friendly implementation and a graphical user interface (GUI) but it also has performance limitations, as will be explained below.

To integrate the RF canceller configuration with the main GNU Radio program, we implemented a customized GNU Radio out-of-tree (OOT) SUB-20 module. Given the relatively stable wireless environment in the ORBIT testbed, the OOT SUB-20 module is implemented with fixed\footnote{Users can change \mytexttt{CAPi} using the SUB-20 \mytexttt{C} program (see Section~\ref{sec:exp-simple}).} $\mytexttt{CAP1} = 16$, $\mytexttt{CAP2} = \mytexttt{CAP3} = 6$, and users can vary the values of \mytexttt{ATT} and \mytexttt{PS} to observe different RF SIC performance. This example FD experiment contains three parts:
\begin{enumerate}[leftmargin=*,label=\arabic*.]
\item \textbf{Data Generation}: The baseband samples encoding raw bits modulated using an PSK scheme (e.g., BPSK and QPSK) are generated using \mytexttt{gen\_data\_psk};
\item \textbf{Data Transmission and RF SIC}: The FD transceiver transmits the modulated samples and receives samples over-the-air using \mytexttt{usrp\_txrx\_psk}. The RF canceller can be configured in \emph{real-time} to observe different RF SIC performance;
\item \textbf{Digital SIC}: The digital SIC is performed offline using \mytexttt{dig\_sic\_on} and the received baseband samples.
\end{enumerate}
Due to the software and timing limitations of GNU Radio, baseband samples are recorded using the file option and digital SIC is performed offline (part 3). We remark that other implementations (e.g., UHD- or FPGA-based) may be able to support digital SIC in real-time.

\begin{figure}[!t]
\centering
\subfloat[]{
\label{fig:exp-psk-node}
\includegraphics[width=0.48\columnwidth]{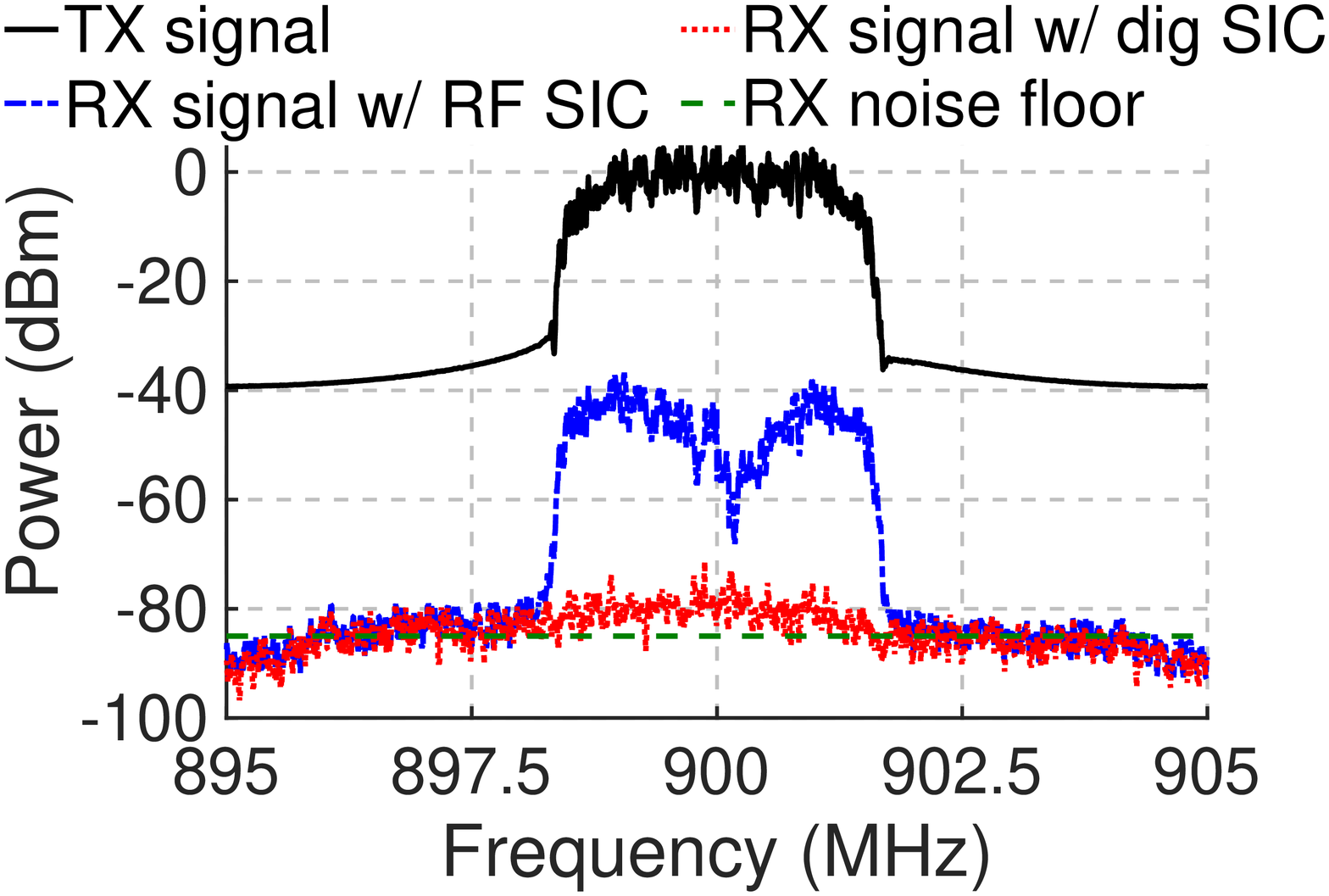}}
\hspace{-12pt} \hfill
\subfloat[]{
\label{fig:exp-psk-link}
\includegraphics[width=0.48\columnwidth]{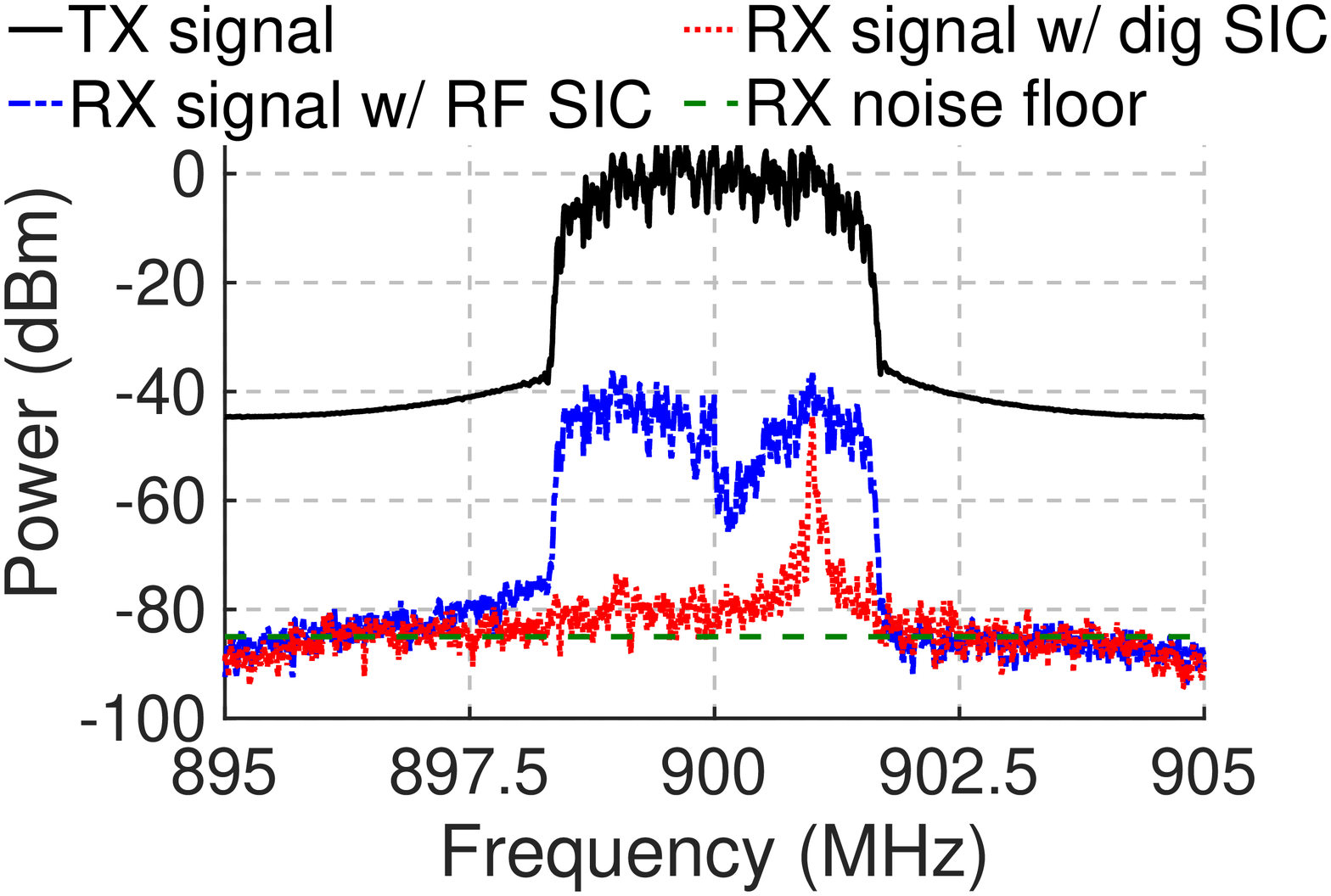}}
\vspace{-0.5\baselineskip}
\caption{Power spectrum of the received signal at the FD transceiver, which transmits a $\SI{2.5}{MHz}$ QPSK signal at $\SI{0}{dBm}$ average TX power level: (a) without the desired signal, (b) with the desired signal.}
\label{fig:exp-psk}
\vspace{-0.5\baselineskip}
\end{figure}

A GNU Radio-based example experiment was demonstrated in~\cite{fd_demo_infocom18} where the FD transceiver (\mytexttt{node11-10}) transmits a $\SI{2.5}{MHz}$ QPSK signal stream with QPSK modulation scheme at $\SI{10}{MHz}$ sampling rate and $\SI{0}{dBm}$ average TX power level. Fig.~\ref{fig:exp-psk}\subref{fig:exp-psk-node} shows the power spectrum of the received signal at the FD transceiver, where $\SI{85}{dB}$ overall SIC is achieved, where $\SI{43}{dB}$ is from the RF domain and $\SI{42}{dB}$ is from the digital domain. The SI signal is canceled to the USRP receiver noise floor at $\SI{0}{dBm}$ TX power. Another ORBIT node (\mytexttt{node13-8}) is then used to serve as a second radio that transmits a single tone with a frequency offset of $\SI{1}{MHz}$. As Fig.~\ref{fig:exp-psk}\subref{fig:exp-psk-link} shows, the desired signal is recoved after the SIC (in both the RF and digital domains) is performed at the FD transceiver.

\section{Other Potential FD Wireless Experiments}
Some potential FD experiments that can be conducted using the presented FD transceiver are listed below:
\begin{itemize}[leftmargin=*]
\item[-] Hands-on experiments with FD wireless on an SDR platform in a teaching/lab course;
\item[-] Studying different RF SIC performance and its relation to the antenna interface response by tuning the RF canceller box (the SUB-20 \mytexttt{C} program or the OOT SUB-20 module);
\item[-] Studying the performance of the digital SIC algorithm by tuning its parameters (digital SIC part of the GNU Radio/UHD program);
\item[-] Development and evaluation of different digital SIC algorithms (digital SIC part of the GNU Radio/UHD program);
\item[-] Incorporation of modulated signals, such as OFDM, with different bandwidth (the GNU Radio/UHD program);
\item[-] Experimentation and evaluation of medium access control (MAC) algorithms in a heterogeneous network with an FD access point/client (e.g., modifying the GNU Radio/UHD program and adding a MAC layer).
\end{itemize}

\section{Conclusion}
\label{sec:conclusion}
In this report, we presented our cross-layered (hardware and software) design and implementation of the first open-access remotely-accessible FD transceiver which is integrated with the ORBIT wireless testbed. An FD transceiver baseline program and an example FD experiment were provided to facilitate the experimentation with the FD transceiver. We discussed other potential FD experiments that can be developed and conducted using the FD transceiver.

Our future work includes the integration of the Gen-2 canceller box in both the ORBIT testbed and the PAWR COSMOS testbed. In particular, we demonstrated the Gen-2 RF canceller in~\cite{fd_demo_infocom17}, which can achieve wideband RF SIC via the technique of frequency-domain equalization. The Gen-2 RF SI canceller implemented on a PCB emulates its RFIC counterpart we presented in~\cite{Zhou_WBSIC_JSSC15}. We plan to install more FD transceivers in the ORBIT and COSMOS testbeds with both Gen-1 and Gen-2 RF canceller boxes. Our future work also includes developing more advanced FD-related software and applications.

\section*{Acknowledgments}
This work was supported in part by NSF grants ECCS-1547406 and CNS-1827923, DARPA RF-FPGA program, DARPA SPAR program, a Qualcomm Innovation Fellowship, Texas Instruments, Intel, and a National Instruments Academic Research Grant. We thank Steven Alfano, Jelena Diakonikolas, Aishwarya Rajen, Jinhui Song, Mingyan Yu for their contributions to various aspects of the project. We thank Ivan Seskar, Jakub Kolodziejski, and Prasanthi Maddala from WINLAB, Rutgers University, for their help on the integration with the ORBIT testbed. We also thank Kira Theuer and Kendall Ruiz from NI and the NI technical support team for their help.

\scriptsize
\bibliographystyle{IEEEtran}
\interlinepenalty=10000
\bibliography{paper_orbit}

\end{document}